\documentclass[a4paper,10pt]{article}
\usepackage{graphicx}
\usepackage{epstopdf}
\usepackage{color}
\usepackage[normalem]{ulem}
\usepackage{dcolumn}

\usepackage{multirow}

\usepackage{amsfonts,amsmath,amssymb}
\usepackage{hyperref}

\newcommand{\f}[2]{\frac{#1}{#2}}
\usepackage{epsfig,multicol}

\newcommand\fverb{\setbox\pippobox=\hbox\bgroup\verb}
\newcommand\fverbit{\egroup\item[\fbox{\unhbox\pippobox}]}

\newbox\pippobox

\begin{document}

\title{Noncommutative Universe and Chameleon Field Dynamics}
\author{Nasim Saba\thanks{Electronic address: n$\_$saba@sbu.ac.ir}\, and\,
        Mehrdad Farhoudi\thanks{Electronic address: m-farhoudi@sbu.ac.ir}\,\,
\\
\small Department of Physics, Shahid Beheshti University, G.C.,
       Evin, Tehran 19839,  Iran}
\date{\small May 12, 2018}
\maketitle
\def\be{\begin{equation}}
\def\ee{\end{equation}}
\def\bea{\begin{eqnarray}}
\def\eea{\end{eqnarray}}
\def\f{\frac}
\def\n{\nonumber}
\def\l{\label}
\def\p{\phi}
\def\o{\over}
\def\R{\rho}
\def\pa{\partial}
\def\om{\omega}
\def\na{\nabla}
\def\P{\Phi}

\begin{abstract}
\noindent
 We consider a noncommutative standard model with a
minimal coupling scalar field and a dynamical deformation between
the canonical momenta of its scale factor and scalar field, and a
chameleon model with a non--minimally coupling scalar field. We
indicate that there is a correspondence between these two models,
more specific, actually between the noncommutative parameter and
the chameleon coupling strength, and also between the matter
density of the chameleon model and the noncommutative geometry. In
addition, the analogy constrains the type of the matter field in
the chameleon model to be nearly a cosmic string--like during the
inflation. Thus, the scenario enables the evolution of the
universe being described by one single scalar field. That is, the
effects of the chameleon and the inflaton can be described by one
single scalar field which plays the role of inflaton in the very
early universe and then, acts as a chameleon field. Moreover, the
proposed correspondence procedure not~only sets some constraints
on the noncommutative parameter and the chameleon coupling
constant, but also nearly specifies functions of the scalar field
and its potential.
\end{abstract}
\noindent
 PACS numbers: 02.40.Gh, 04.50.Kd, 98.80.-k, 98.80.Cq\\
 Keywords: Noncommutative Geometry; Chameleon
          Cosmology; Inflationary Universe.

\bigskip

\section{Introduction}
\indent

Alternative theories of gravitation have almost a long history and
different motivations\rlap,$^{1}$\footnotetext[1]{See, e.g.,
Refs.~\cite{farc}--\cite{farb} and references therein.}\
 among which the scalar--tensor theories of
gravitation, that extend general relativity by introducing a new
degree of freedom, become one of the most popular alternatives to
the Einstein gravitational theory, see, e.g.,
Refs.~\cite{fujji}--\cite{capozziello}. However, it is obvious
that any proposed theory in this issue will be valid if it
satisfies both the current astronomical observations and the
laboratory experiments. In this regard, although the
scalar--tensor theories are claimed to be successful in explaining
the present accelerated expansion of the universe (that has been
declared via analyzing the various observed cosmological
data~\cite{1}--\cite{p152}), yet there are some problems in these
theories in describing the results of the solar system tests of
gravity. This shortcoming is mostly caused due to coupling between
the scalar and the matter fields. Actually, due to such a
coupling, the detection of a fifth force and also violation of the
equivalence principle (EP) are expected~\cite{fi,euc} that are in
contrast with the results of the solar system tests of
gravity~\cite{iess}--\cite{dec}. In this respect, the chameleon
model~\cite{weltman1}--\cite{t7} (originally proposed by Weltman
and Khoury in 2004~\cite{weltman1,weltman2}) is a scalar--tensor
theory that, due to interaction of the chameleon field with an
ambient matter field, invokes a screening
mechanism~\cite{screen1}--\cite{Burrage2017} which uses
non--linear effects to suppress deviations from general relativity
in the solar system while allowing them to be relevant on larger,
i.e. cosmological, scales. Hence, it remains consistent with the
tests of gravity both on the terrestrial and the solar system
scales with no need to tune free parameters (masses and couplings)
to evade solar system tests. In fact, the scalar field in the
chameleon cosmology, similar to the quintessence
models~\cite{peebles}--\cite{bamba}, provides a dynamical
alternative to the static cosmological constant. Due to the
coupling of the scalar and the matter fields with the
gravitational strength, the chameleon field acquires a
density--dependent mass and consequently, its property changes
depending on the environmental situations. Because of such a
dependence, its mass is very light in the cosmological scales,
wherein it may play the role of dark energy and causes the cosmic
late time acceleration. Although, in the regions of high density,
such as on the earth, the chameleon field acquires a large mass
that makes its effects being short--ranged and hence, becomes
invisible in search for the EP--violation and fifth force in the
current experimental and observational tests.

However, in Refs.~\cite{wang2,khoury} by proving two theorems, it
has been claimed that the effect of chameleon--like scalar fields
is negligible on density perturbations considering the linear
scales and also, its influence may~not be regarded for the
observed cosmic acceleration except as some form of dark energy.
Also, in Ref.~\cite{saba}, we considered a coupling between the
chameleon field and an unknown matter scalar field, and analyzed a
possibility of an influence of the chameleon field on a
cosmological acceleration of the universe inflation. In the
context of the slow--roll approximations, by employing the
potential usually used in this issue in the literature, we
evaluated the number of e--folding of the model in order to verify
its viability during the inflation. We examined the model for
different ranges of the free parameters, however, the results of
analysis showed that there is~not much chance of having the viable
chameleonic universe inflation. That is, as expected for the
extreme case, the exponential term (entered due to the conformal
factor describing the interaction of the chameleon field with an
ambient matter) in the resulted effective potential cannot inflate
unless one fixes the value of the scalar, which in turn means that
the matter density must be constant.

Nevertheless, in probing the role of the chameleon field on the
cosmological scales, we noticed it has been shown~\cite{ivanov}
that, although the cosmological factor would be in principle
constant during the last Hubble time, but this point does~not keep
the chameleon field being accounted for a late--time acceleration
of the universe expansion. Also, although the influence of the
chameleon field on the universe inflation is insignificant, but it
has been indicated~\cite{ivanov2} that the cosmological constant
as well as dark energy density might be induced by torsion in the
Einstein--Cartan gravitational theory. Indeed, as the torsion is a
natural geometrical quantity in addition to the metric tensor, it
supplies a robust geometrical background in the origin of those.
Moreover, as mentioned in Ref.~\cite{saba}, even in the limiting
case, where the coupling strength in the exponential term of the
conformal factor tends to zero (i.e., reducing to the minimal
coupling case), still one can trace its effect in the
corresponding Klein--Gordon equation. That is, it looks as if the
ambient matter fluid somehow acts like an extra field and hence,
it would have some effects during the chameleonic inflation.

In this sense, encouraged by these issues, we have proposed in
Ref.~\cite{saba} that if the chameleon model, through some
mechanism, reduces to the standard inflationary model during the
inflation, then it may cover the whole era of the universe from
the inflation up to the late time. In this respect, the task of
the present work is to investigate such a scenario at the very
early universe.

Besides, it should be mentioned that in Refs.~\cite{h1,ch}, via a
supersymmetric potential introduced in Ref.~\cite{kklt}, a
chameleon scenario has been embedded within the string
compactifications, and hence, it has been shown that the volume
modulus of the compactification can act as a chameleon field. The
late time investigation of this scenario has been presented in
Ref.~\cite{h1}, while Ref.~\cite{ch} describes it during the
inflation. Hence, it has been indicated that in order to cover the
cosmology of both the late time and the very early universe, there
exists a superpotential consisting of two pieces. One piece drives
inflation in the very early universe, and the other one is
responsible for the chameleon screening at the late times.

Moreover, in recent years, some attempts have been devoted to
search for an inflationary solution in the context of the string
theory~\cite{vafa}--\cite{mota} owing to the fact that the
inflation would typically occur at energies near the string scale.
On the other hand, the noncommutativity of the spacetime (first
introduced by Snyder~\cite{a} and nowadays known as noncommutative
geometry~\cite{Connes94,Hinch}) has also attracted some attentions
due to the strong motivation of the string and M--theories, see,
e.g., Refs.~\cite{d}--\cite{u}. The noncommutativity effects are
significant at the Planck scale where quantum gravity effects are
prominent. Indeed, during the inflation (that is a period of
rapidly accelerating expansion soon after the bing
bang~\cite{guth}--\cite{Campo}), the energy scale of the universe
is extremely high, hence, it is reasonable to take into
consideration the noncommutativity effects, as for instance, such
effects have been considered in some other contexts, see, e.g.,
Refs.~\cite{k}--\cite{ee}. In addition, some efforts have been
devoted to explain the standard model of particle physics in the
framework of noncommutative spectral geometry and the cosmological
consequences of such an approach~\cite{mairi,mairi0}.

In this work, we study the effects of noncommutativity in the
standard scalar--tensor model in which gravity is minimally
coupled with a dynamical scalar field, and then, compare it with
the chameleon mechanism while investigating an analogy between
these two models. Actually, as a complementary to our previous
work~\cite{saba}, such an analogy may provide a scenario to overcome the
issue of the chameleon model during the inflationary epoch.

The work is organized as follows. In the next section, we briefly
derive the cosmological equations of the standard scalar--tensor
model in the presence of a specified noncommutativity parameter.
In Sect.~III, we revisit the chameleon model and concisely derive
the equations governing its evolution. However, we furnish these
two sections with almost different expression and in a way to
provide our requirements for the proposal of the work. The main
purpose of this study is to compare these two models, where this
goal is performed in Sect.~IV. Finally, we conclude the summary of
the results in the last section. Throughout the work, we take the
signature of spacetime to be $(-,+,+,+)$, $c=1=\hbar$ and the
reduced Planck mass $M_{\rm Pl}\equiv(8\pi G)^{-1/2}$.

\section{Noncommutative Standard Scalar--Tensor Model}
\indent

In this section, we outline the standard scalar--tensor model with
a minimal coupling between gravity and the scalar field in the
context of noncommutativity described by the action
\begin{equation}
S=\int{\rm
d}^{4}x\sqrt{-g}\left(\frac{M^{2}_{\rm
Pl}R}{2}\!-\!\frac{1}{2}\partial_{\mu}\phi \partial^{\mu}\phi-V^{{[\rm NC]}}(\phi)\right),\label{eq1}
\end{equation}
where $R$ is the Ricci scalar constructed from the metric
$g_{\mu\nu}$, $g$ is the determinant of the metric and the
lowercase Greek indices run from zero to three. Also, $\phi$ is a
scalar field and $V^{{[\rm NC]}}(\phi)$ is a potential for this model.

In addition, we consider a homogeneous scalar field, that is
$\phi=\phi(t)$, and also assume that the background spacetime is
spatially flat, homogeneous and isotropic given by the
Friedmann--Lema\^{\i}tre--Robertson--Walker ({\bf FLRW}) metric
\begin{equation}
ds^{2}=-N^{2}(t)dt^{2}+a^{2}(t)\left(dx^{2}+dy^{2}+dz^{2}\right),\label{frw}
\end{equation}
where $N(t)$ is a lapse function, $a(t)$ is the scale factor
describing the cosmological expansion, and $t$ is the cosmic time.
It is straightforward to show that the Hamiltonian of the model is
\begin{equation}
\mathcal{H}_0=-\frac{1}{12M^{2}_{\rm
Pl}}Na^{-1}P_{a}^{2}+\frac{1}{2}Na^{-3}P_{\phi}^{2}+Na^{3}V^{{[\rm NC]}}(\phi),\label{H0}
\end{equation}
where $P_{a}$ and $P_{\phi}$ are the conjugate momenta associated
to the scale factor and the scalar field, respectively. However,
as the conjugate momentum of $N(t)$ vanishes, one has to add it as
a constraint to the Hamiltonian and hence, the corresponding Dirac
Hamiltonian is
\begin{equation}\label{DiracH}
{\cal H}={\cal H}_{0}+\lambda P_N,
\end{equation}
where $\lambda$ is a Lagrange multiplier.

First, let us derive the equations of motion corresponding to
the commutative phase space coordinates, i.e. $\{a,\phi, N;P_a,P_\phi, P_N\}$,
in which the ordinary phase space structure is described by the
usual Poisson brackets
\begin{equation}\label{NotDeformed}
\{a,P_{a}\}=1=\{\phi,P_{\phi}\} \qquad\qquad {\rm and}\qquad\qquad
\{N,P_N\}=1,
\end{equation}
while the other brackets vanish. Applying these brackets and using
Hamiltonian (\ref{DiracH}), the equations of motion are obtained
as
\begin{eqnarray}
\dot{a}\!\!\! & = & \!\!\!\{a,{\mathcal{H}}\}=-\frac{1}{6M^{2}_{\rm Pl}}Na^{-1}P_{a},\label{diff.eq1}\\
\dot{P}_{a}\!\!\! & = &
\!\!\!\{P_{a},{\mathcal{H}}\}=-\frac{1}{12M^{2}_{\rm
Pl}}Na^{-2}P_{a}^{2}
 +\frac{3}{2}Na^{-4}P_{\phi}^{2}
  - 3Na^{2}V^{{[\rm NC]}}(\phi),\label{diff.eq2}\\
\dot{\phi}\!\!\! & = & \!\!\!\{\phi,{\mathcal{H}}\}=Na^{-3}P_{\phi},\label{diff.eq3}\\
\dot{P}_{\phi}\!\!\! & = & \!\!\!\{P_{\phi},{\mathcal{H}}\}=-Na^{3}V'^{{[\rm NC]}}(\phi),\label{diff.eq4}\\
\dot{N}\!\!\! & = & \!\!\!\{N,{\mathcal{H}}\}=\lambda,\label{diff.eq5}\\
\dot{P}_{N}\!\!\! & = &
\!\!\!\{P_{N},{\mathcal{H}}\}\!=\!\frac{1}{12M^{2}_{\rm
Pl}}a^{-1}P^{2}_{a}\!\!-\!
 \frac{1}{2}a^{-3}P_{\phi}^{2}\!\!-\!a^{3}V^{{[\rm NC]}}(\phi),\,\label{diff.eq6}
\end{eqnarray}
where dot and prime denote the derivative with respect to the
cosmic time and the scalar field, respectively. Furthermore, to
satisfy the constraint $P_N =0$ at all times, the secondary
constraint $\dot{P}_N= 0$ should also be satisfied.

Then, in order to investigate the proposed effect of the
noncommutativity, inspired by the motivations mentioned in
Ref.~\cite{MF16,RSFM}, we leave all the Poisson brackets for the
noncommutative variables (shown with the tilde sign) unchanged,
except the following plausible dynamical deformation between the
canonical conjugate momenta of the scale factor and of the scalar
field. That is, we deliberately choose
\begin{equation}\label{deformed}
\{\tilde{P}_{\tilde{a}},\tilde{P}_{\tilde{\phi}}\}=\ell M^{3}_{\rm Pl}e^{\ell\tilde{\phi}},
\end{equation}
where $\ell$ is a constant length indicator parameter that can
present and trace the quantum behaviors (see, e.g.,
Ref.~\cite{MF16}), and if it vanishes, the standard (classical)
counterpart relations will be recovered. Of course this choice,
that satisfies the dimensionality
requirements\rlap,$^{1}$\footnotetext[1]{It is easy to show that
the dimensions of $P_a$, $P_\phi$, $\phi$, $\ell$ and consequently
$\{P_a,P_\phi\}$ are $L_{\rm P}^{-3}$, $L_{\rm P}^{-2}$, $L_{\rm
P}^{-1}$, $L_{\rm P}$ and $L_{\rm P}^{-2}$, respectively, where
$L_{\rm P}=\sqrt{\hbar G/c^3}$ is the Planck length, however we
have used the units in which $c=1=\hbar$.}\
 is restrictive, but we limit
ourselves to it for reasons of simplicity which easily leads to
the desired results (see Sect.~IV).

In this case, the minimally deformed (noncommutative) version of
the Dirac Hamiltonian is achieved by replacing the untilde
variables with the tilde ones in relations (\ref{H0}) and
(\ref{DiracH}), namely
\begin{equation}\label{nonDiracH}
\mathcal{\tilde{H}}=-\frac{1}{12M^{2}_{\rm
Pl}}\tilde{N}\tilde{a}^{-1}\tilde{P}^{2}_{\tilde{a}}+\frac{1}{2}\tilde{N}
\tilde{a}^{-3}\tilde{P}^{2}_{\tilde{\phi}}+\tilde{N}\tilde{a}^{3}V^{{[\rm NC]}}(\tilde{\phi})
 +\tilde{\lambda}\tilde{P}_{\tilde{N}},
\end{equation}
where we have kept the noncommutative Hamiltonian with the same
functional form as in (\ref{H0}). However, from now on, for
simplicity and less confusion, we drop the tilde sign for the
noncommutative variables, and also work in the comoving gauge,
namely, we set $N(t)=1$.

At this stage, by applying the new bracket (\ref{deformed}), the
equations of motion associated to the new variables $a$, $\phi$,
$N$ and $P_{N}$ are left unchanged while the equations of motion
for the new momenta sectors are deformed as
\begin{equation}\label{diff.eq-prime2}
\dot{P}_{a} =\{P_{a},{\mathcal{H}}\} = -\frac{1}{12M^{2}_{\rm
Pl}}a^{-2}P_{a}^{2}+ \frac{3}{2}a^{-4}P_{\phi}^{2}-
3a^{2}V^{{[\rm NC]}}(\phi)+\ell a^{-3}M^{3}_{\rm Pl}P_{\phi}e^{\ell\phi},
\end{equation}
\begin{equation}
\dot{P}_{\phi}=\{P_{\phi},{\mathcal{H}}\}=-a^{3}V'^{{[\rm NC]}}(\phi)+
\frac{1}{6} \ell a^{-1}M_{\rm
Pl}P_{a}e^{\ell\phi}.\label{diff.eq-prime4}
\end{equation}
As mentioned, the usual brackets are restored by setting $\ell=0$.
With the aid of Eqs. (\ref{diff.eq1}), (\ref{diff.eq3}),
(\ref{diff.eq6}) and also (\ref{diff.eq-prime2}) and
(\ref{diff.eq-prime4}), after performing some manipulations, the
noncommutative equations of motion are achieved as
\begin{equation}\label{fn1}
3M^{2}_{\rm
Pl}H^2=\frac{1}{2}\dot{\phi}^{2}+V^{{[\rm NC]}}(\phi)=\rho^{{[\rm NC]}}_{\phi}\equiv\rho_{\rm tot}
,
\end{equation}
\begin{equation}\label{fn2}
{M^{2}_{\rm Pl}}\left(2\frac{\ddot{a}}{a}+H^{2}\right)=
-\left(\frac{1}{2} \dot{\phi}^{2}-V^{{[\rm NC]}}(\phi)+\frac{1}{3}\ell
a^{-2}\dot{\phi}M^{3}_{\rm Pl}e^{\ell\phi}\right)\equiv
-\left(p^{{[\rm NC]}}_{\phi}+p_{\ell}\right)\equiv - p_{\rm tot},
\end{equation}
\begin{equation}\label{nphi}
\ddot{\phi}+3H\dot{\phi}+V'^{{[\rm NC]}}(\phi)=-\ell Ha^{-2}M^{3}_{\rm
Pl}e^{\ell\phi}\equiv Q/\dot{\phi},
\end{equation}
where $H(t)\equiv\dot{a}/a$ is the Hubble expansion rate of the
universe and we have defined a deformed pressure density as
\begin{equation}\label{deformed.p}
p_{\ell}\equiv \frac{1}{3}\ell a^{-2}\dot{\phi}M^{3}_{\rm
Pl}e^{\ell\phi}
\end{equation}
and the $Q$ term as
\begin{equation}\label{definedQ}
Q\equiv -\ell Ha^{-2}\dot{\phi}M^{3}_{\rm Pl}e^{\ell\phi}.
\end{equation}
Moreover, the time derivation of the Hubble parameter is
\begin{equation}\label{hdot}
-2M^{2}_{\rm
Pl}\dot{H}=\dot{\phi}^{2}+\frac{1}{3}\ell
a^{-2}\dot{\phi}M^{3}_{\rm Pl}e^{\ell\phi}.
\end{equation}

As it is obvious, by taking the noncommutativity effect into
account, the Klein--Gordon equation and the total pressure density
have been obtained different from the corresponding usual
commutative ones. However, if one sets $\ell= 0$, then each
equation will have the same form as its corresponding commutative
one that usually is applied for the inflationary scenario. It is
worth noting that the noncommutative theory has led to an
additional force, however this assertion cannot be expressed only
from the background equations alone and a perturbative analysis
might be necessary. Nevertheless, in a similar manner of
methodology employed in Ref.~\cite{MF21} and since a conservative
force is obtained from the derivative of a potential, the right
hand side of Eq. (\ref{nphi}) can be interpreted as an additional
force. Besides, the term $3H\dot{\phi}$ in the left hand side of
Eq. (\ref{nphi}) is a dissipative force. Also, definitions
$p_{\ell}$ and $Q$ (relations (\ref{deformed.p}) and
(\ref{definedQ})) yield
\begin{eqnarray}\label{conservPl}
3Hp_{\rm \ell}=-Q.
\end{eqnarray}

On the other hand, as the conservation equation of the model is
\begin{equation}\label{conservNC}
\dot{\rho}_{\rm tot}+3H\left(\rho_{\rm
tot}+p_{\rm tot}\right)=0,
\end{equation}
with the aid of relation (\ref{conservPl}), it leads to
\begin{eqnarray}\label{conservphiQ}
\dot{\rho}^{{[\rm NC]}}_{\phi}+3H(\rho^{{[\rm NC]}} _{\phi}+p^{{[\rm NC]}} _{\phi})= Q.
\end{eqnarray}
That is, while the effects of noncommutativity manifest itself in
the pressure term $p_{\rm \ell}$, the results dictate that, even though the total
energy density is conserved, but the corresponding one for the
scalar field and the one for the noncommutativity part (i.e., Eqs.
(\ref{conservphiQ}) and (\ref{conservPl}), respectively) are~not
separately conserved. Indeed, their conservation equations are~not
independent and the $Q$ term stands as an interacting term among
them. Such a similar interacting term has been employed in the
literature, see, e.g., Refs.~\cite{Movahed,Bahrehbakhsh-2013}.
\section{Chameleon Model}
\indent

In this section, we give a brief review of the chameleon scalar
field model, and derive the equations governing the cosmological
evolution of the universe. In this regard, we start with the
well--known action that governs the dynamics of the chameleon
scalar field model in 4--dimensions, i.e.$^{1}$\footnotetext[1]{In
the next section, we discuss on equality and/or differentness of
those parameters that appear in this section and the previous
one.}
\begin{equation}\label{Ch.action}
S=\int{\rm d}^{4}x\sqrt{-g}\left[\frac{M^{2}_{\rm
Pl}R}{2}-\frac{1}{2}\partial_{\mu}\phi
\partial^{\mu}\phi-V^{{[\rm CH]}}(\phi)\right]+\sum_{i}\int{\rm d}^{4}x\sqrt{-{\tilde g}^{(i)}}{L_{\rm m}^{(i)}}
 \left({\psi}^{(i)},{\tilde g}^{(i)}_{\mu\nu}\right),
\end{equation}
where $V^{{[\rm CH]}}(\phi)$ is a potential in the chameleon model, $L_{\rm m}^{(i)}$'s~are the
Lagrangians of the matter fields and $\psi^{(i)}$'s~are various
matter scalar fields. Also, $\tilde g_{\mu\nu}^{(i)}$'s are the
matter field metrics that are conformally related to the Einstein
frame metric $g_{\mu\nu}$ via
\begin{equation}
{\tilde g}^{(i)}_{\mu \nu}={e}^{ {2\frac {{\beta} _{i}\phi}{M_{\rm Pl}}}}{g}_{\mu \nu},
\end{equation}
where $\beta_{i}$'s are dimensionless constants representing
different non--minimal coupling constants between the scalar field
and each one of the matter species as the strength of the matter
couplings. However, for simplicity, we just focus on a single
matter component, and henceforth, we drop the index $i$. Besides,
the potential shown in action (\ref{Ch.action}) is an arbitrary
function, whereas the chameleon effect is a mechanism to hide the
fifth force mediated by the scalar field, for which certain
conditions on potential are needed. Nevertheless in the following
analysis, in order to match the two models, we end up on some
necessary conditions on the potential that are mostly compatible
with the chameleon model.

Moreover, we consider the FLRW metric (\ref{frw}) in the comoving
gauge and a homogeneous scalar field. In this case, the
corresponding metric in the Jordan frame is
\begin{equation}\label{metric2}
d\tilde{s}^{2}=-{e}^{ {2\frac {{\beta}\phi}{M_{\rm Pl}}}}dt^{2}+\tilde{a}^{2}(t)\left(dx^{2}+dy^{2}+dz^{2}\right),
\end{equation}
where $\tilde{a}(t)$ is the scale factor in this frame, i.e.
$\tilde{a}(t)\equiv a(t) \exp({\beta\phi/M_{\rm Pl}})$. Varying
action (\ref{Ch.action}) with respect to the scalar field yields
the field equation of motion
\begin{equation}\label{box}
\Box\phi=V'^{{[\rm CH]}}(\phi)-\frac{{\beta}}{M_{\rm
Pl}}e^{4\frac{{\beta}\phi}{{M}_{\rm Pl}}}{\tilde
g}^{\mu\nu}{\tilde T}_{\mu\nu},
\end{equation}
where $\Box\equiv\nabla^{\alpha}\nabla_{\alpha}$ corresponding to
the metric $g_{\mu\nu}$ and $\tilde T_{\mu\nu}=-(2/\sqrt{-\tilde
{g}})\delta (\sqrt{-\tilde {g}}L_{\rm m})/\delta {\tilde
g}^{\mu\nu}$ is the energy--momentum tensor that is conserved in
the Jordan frame, i.e. $\tilde\nabla_{\mu} {\tilde T^{\mu\nu}}=0$.
We also assume the matter field as a perfect fluid with the
equation of state $\tilde p=w\tilde \rho$, where, in the FLRW
background, one has
\begin{equation}\label{trace}
{\tilde g}^{\mu\nu}\tilde T_{\mu\nu}=-(1-3w)\tilde \rho
\end{equation}
with $\tilde\rho$ as the matter density in the Jordan frame.

Since $\tilde\rho$ is~not conserved in the Einstein frame, we
propose to have a conserved matter density that is independent of
$\phi$ and obeys the relation
\begin{equation}\label{cont1}
\rho=\rho_{0} a^{-3(1+w)}
\end{equation}
in the Einstein frame, i.e. with the fluid equation
$\dot{\rho}+3H(1+w)\rho=0$. In relation (\ref{cont1}), $\rho_{0}$
is a constant for the matter density that acts as a cosmological
constant with $w=-1$. For this purpose, as the continuity equation
for $\tilde\rho$ in the Jordan frame is
\begin{equation}\label{cont2}
\dot{\tilde{\rho}}+3\frac{\dot{\tilde{a}}}{\tilde{a}}(1+w)\tilde{\rho}=0,
\end{equation}
i.e. $\left(\tilde{a}^{3(1+w)}\tilde{\rho}\right)_{,0}=0$, hence
in order to have relation (\ref{cont1}), one gets
\begin{equation}\label{rhoo}
\tilde {\rho}= e^{-3(1+w)\frac{{\beta}\phi}{M_{\rm Pl}}}\rho.
\end{equation}
Therefore, by relations (\ref{trace}) and (\ref{rhoo}), Eq.
(\ref{box}) reads
\begin{equation}\label{field}
\Box\phi=V'^{{[\rm CH]}}(\phi)+\rho(1-3w)\frac{\beta}{{M}_{\rm Pl}}
e^{(1-3w)\frac{{\beta}\phi}{M_{\rm Pl}}}\equiv V'_{\rm eff}(\phi).
\end{equation}
This implies that the dynamic of the scalar field is~not ruled
only by the self--interacting potential $V^{{[\rm CH]}}(\phi)$, but it is
actually governed by an effective potential defined as
\begin{equation}\label{eff}
{V}_{\rm eff}(\phi)\equiv V^{{[\rm CH]}}(\phi)+\rho
e^{(1-3w)\frac{\beta\phi}{M_{\rm Pl}}},
\end{equation}
which depends on the background matter density $\rho$ of the
environment. Consequently, the value of $\phi$ at the minimum of
$V_{\rm eff}$ and the mass fluctuation about the minimum depend on
the matter density which can give a chance to the chameleon field
to be hidden from local experiments.

Now, by the FLRW metric (\ref{frw}), the field equation
(\ref{field}) reads
\begin{equation}\label{phi}
\ddot{\phi}+3H\dot{\phi}+ V'^{{[\rm CH]}}(\phi)=-\rho (1-3w)\frac{\beta}{M_{\rm
Pl}}e^{(1-3w)\frac{\beta\phi}{M_{\rm Pl}}}\equiv X/\dot{\phi},
\end{equation}
where
\begin{equation}\label{definedX}
X\equiv -\rho(1-3w)\frac{\beta}{M_{\rm Pl}}\dot{\phi}
e^{\left(1-3w\right)\frac{\beta\phi}{M_{\rm Pl}}}.
\end{equation}
In addition, the variation of action (\ref{Ch.action}) with
respect to the metric tensor $g_{\mu \nu}$ associated to the
Einstein frame gives the field equations
\begin{equation}\label{G}
{G_{\mu \nu }}=\frac{1}{{M_{Pl}^2}}\left( {T_{\mu \nu }^{[\beta]}
+ T_{\mu \nu }^{[\phi]} } \right),
\end{equation}
where $ T_{\mu \nu }^{[\beta]}$ and $T_{\mu \nu }^{[\phi]} $ are
the energy--momentum tensors of the coupling term and the scalar
field, respectively, defined as
\begin{equation}\label{pphiii}
T_{\mu \nu }^{[\phi]}  \equiv  - \frac{1}{2}{g_{\mu \nu
}}{\partial ^\alpha }\phi {\partial _\alpha }\phi  - {g_{\mu \nu
}}V^{{[\rm CH]}}\left( \phi  \right) + {\partial _\mu }\phi {\partial _\nu
}\phi,
\end{equation}
\begin{equation}\label{te}
 T^{[\beta]}_{\mu\nu}\equiv -(2/\sqrt{-g})\delta
{\cal L}_{\rm m}/\delta {g}^{\mu\nu}.
\end{equation}
One can easily verify that
\begin{equation}
 T^{[\beta]}_{\mu\nu}={e}^{ {2\frac {{\beta}\phi}{M_{\rm Pl}}}}{\tilde T}_{\mu\nu},
\end{equation}
then by the conservation of ${\tilde T}_{\mu\nu}$ in the Jordan
frame, we obtain
\begin{equation}\label{nab}
\nabla_{\mu}{T^{[\beta]}}^{0\mu}=X.
\end{equation}
That is, ${T^{[\beta]}}^{0\mu}$ is~not conserved in the Einstein
frame and the geodesic equation is affected by the $X$ term that
can be interpreted as an additional force. Also, the
Friedmann--like equations for the model in the context of the
perfect fluid can be obtained as
\begin{equation}\label{friedman}
3M^{2}_{\rm Pl}H^2= \frac{1}{2}\dot\phi^{2}+V^{{[\rm CH]}}(\phi)+\rho
e^{(1-3w)\frac{\beta\phi}{M_{\rm
Pl}}}\equiv \rho^{{[\rm CH]}}_{\phi}+\rho_{\beta}\equiv\rho_{\rm tot}
\end{equation}
and
\begin{equation}\label{fried2}
{M^{2}_{\rm Pl}}\left(2\frac{\ddot{a}}{a}+H^{2}\right)=
-\left(\frac{1}{2} \dot{\phi}^{2}-V^{{[\rm CH]}}(\phi)+w\rho
e^{\left(1-3w\right)\frac{\beta\phi}{M_{\rm Pl}}}\right)\equiv
-\left(p^{{[\rm CH]}}_{\phi}+p_{\beta}\right)\equiv -p_{\rm tot},
\end{equation}
where we have defined a coupled mass density and a coupled
pressure density
\begin{equation}\label{couped.rho.p}
\rho_{\beta}\equiv \rho e^{(1-3w)\frac{\beta\phi}{M_{\rm Pl}}}
\qquad\qquad {\rm and}\qquad\qquad
 p_{\beta}\equiv w\rho
e^{\left(1-3w\right)\frac{\beta\phi}{M_{\rm Pl}}},
\end{equation}
that has the same equation of state as the matter field. Also, the
time derivative of the Hubble parameter in this case is
\begin{equation}\label{hdot2}
-2M^{2}_{\rm Pl}\dot{H}= \dot\phi^{2}+\left(1+w\right)\rho
e^{(1-3w)\frac{\beta\phi}{M_{\rm
Pl}}}.
\end{equation}
Therefore, by taking the non--minimal coupling effect into
account, the Klein--Gordon equation, the total mass and the total
pressure densities have been achieved different from the
corresponding ones without coupling. However, if one sets $\beta=
0$, then each equation will have the same form as its
corresponding standard one.

Moreover, definitions (\ref{definedX}) and (\ref{couped.rho.p}),
with the fluid equation for the matter density
$\rho$, give
\begin{eqnarray}\label{conservPbeta}
\dot{\rho}_{\beta}+3H(\rho _{\beta}+p _{\beta})=-X,
\end{eqnarray}
that is consistent with Eq. (\ref{nab}). On the other hand, as
the conservation equation of this model is also the same as Eq.
(\ref{conservNC}), Eq. (\ref{conservPbeta}) leads to
\begin{eqnarray}\label{conservphiX}
\dot{\rho}^{{[\rm CH]}}_{\phi}+3H(\rho^{{[\rm CH]}}_{\phi}+p^{{[\rm CH]}}_{\phi})= X.
\end{eqnarray}
That is, the effects of non--minimal coupling manifest itself in
both the energy and the pressure densities (i.e., $\rho _{\beta}$
and $p _{\beta}$). Furthermore, the results show that the total
energy density is conserved however, it is~not conserved
separately for the scalar field and its coupling to the matter
field. In another word, their conservation equations are~not
independent and the $X$ term, that stands as an interacting term
among them, manifests itself as a deviation term into the geodesic
equation and are interpreted as an additional force.
\section{Analogy of the Models}
\indent

In this section, as we are interested in having one scalar field
which plays the role of both inflaton and chameleon, hence, we
have considered the scalar field to be the same in the
noncommutative standard and the chameleon models. On the other
hand, since the Hubble constant is an observational quantity, we
have preferred to choose the Hubble parameter also to be the same
in both models\rlap.$^{1}$\footnotetext[1]{However, we have also
examined having different Hubble parameters between two models,
but from mathematical point of view, this choice was~not
fruitful.}\
 Nevertheless, in this case, it is~not necessary that the
corresponding scale factors being exactly the same, however those
are proportional to each other, namely
\begin{equation}\label{scalefactor}
a_{_{[\rm CH]}}(t)=A\, a_{_{[\rm NC]}}(t),
\end{equation}
where $A$ is a constant, and $a_{_{[\rm CH]}}$ and $a_{_{[\rm
NC]}}$ are the scale factors in the chameleon model and in the
noncommutative standard one, respectively, while we have shown
those without any distinguished label in the previous sections.
Moreover, it is~not also necessary that potentials being the same
in the both models, as we have considered them to be different
from the beginning.

Now, we compare the behavior of the other parameters for any
probable correspondence between the models during the inflation
where the Hubble parameter is {\it nearly}
constant\rlap.$^{1}$\footnotetext[1]{A constant Hubble parameter (i.e., say $H_{\rm c}$)
gives $a_{_{[\rm NC]}}(t)={a_{\rm 0}}_{_{[\rm NC]}}e^{H_{\rm c}t}$
and $a_{_{[\rm CH]}}(t)={a_{\rm 0}}_{_{[\rm CH]}}e^{H_{\rm c}t}$ where
${a_{\rm 0}}_{_{[\rm NC]}}$ and ${a_{\rm 0}}_{_{[\rm CH]}}$ are
initial constants for the noncommutative and the chameleon models,
respectively.}\
 In this respect, a direct match of the
corresponding Friedmann Eqs. (\ref{fn1}) and (\ref{friedman})
leads to
\begin{equation}\label{sim0}
V^{[\rm NC]}(\phi)-V^{[\rm CH]}(\phi)=\rho
e^{(1-3w)\frac{\beta\phi}{M_{\rm Pl}}},
\end{equation}
for any $w$ and $\beta$. To proceed further, we have presumed the
following considerations. It is obvious that, in the onset of the
inflation when the matter density of the universe is extremely
high, the effect of coupling term being inefficient in the
effective potential of the chameleon model, and it cannot inflate
unless the matter density approaches a constant value with $w=-1$
(i.e., a cosmological constant). Besides, when in addition
$\beta=0$, the chameleon model reduces to the standard
inflationary model, and in this case, the right hand side of Eq.
(\ref{sim0}), by relation (\ref{cont1}), is $\rho_{0}$. On the
other hand, as it is unlikely that the constant values $\beta$ and
$w$ being appeared in the functionality of the potentials, thus we
likely assume
\begin{equation}\label{simm0}
V^{[\rm NC]}(\phi)-V^{[\rm CH]}(\phi)=\rho_{0}
\end{equation}
to be valid for any $w$ and $\beta$ values. Hence, in turn, one
has
\begin{equation}
 V'^{[\rm NC]}=V'^{[\rm CH]}\equiv V'.
\end{equation}

In this situation, by matching the Klein--Gordon equations of
motion (\ref{nphi}) and (\ref{phi}), we plausibly obtain
\begin{equation}\label{sim1}
\ell \longleftrightarrow (1-3w)\frac{\beta}{M_{\rm Pl}}
\end{equation}
and
\begin{equation}\label{sim2}
\ell H {a^{-2}_{_{[\rm NC]}}}M^{3}_{\rm
Pl}\longleftrightarrow \rho (1-3w)\frac{\beta}{M_{\rm Pl}}.
\end{equation}
Relation (\ref{sim1}) implies that the noncommutative
parameter $\ell$ in the standard model is related to $\beta$ as
the coupling strength of the scalar field to the matter field in
the chameleon model. Inserting this relation into relation
(\ref{sim2}) also leads to
\begin{equation}\label{sim3}
H {a^{-2}_{_{[\rm NC]}}}M^{3}_{\rm Pl} \longleftrightarrow \rho,
\end{equation}
that indicates the matter density of the environment in the
chameleon model also relates to influence of geometry in the
noncommutative standard model. Besides, from relation
(\ref{sim0}), one has
\begin{equation}\label{sim30}
\rho\, e^{\ell\phi}=\rho_{\rm 0},
\end{equation}
that in turn, by relation (\ref{cont1}), gives
\begin{equation}\label{sim300}
a_{_{[\rm CH]}}= e^{\ell\phi/ 3(1+w)}.
\end{equation}

The comparison looks enough if one also matches Eqs. (\ref{fn2})
and (\ref{fried2}) (or equivalently, Eqs. (\ref{hdot}) and
(\ref{hdot2})) that, by relations (\ref{sim1}) and (\ref{sim3}), lead to
\begin{equation}\label{sim4}
\dot{\phi}=\frac{3(1+w)H}{\ell}.
\end{equation}
This result can also be obtained by differentiating relation
(\ref{sim30}) with respect to time while using relation
(\ref{cont1}). Besides, relations (\ref{sim1}) and
(\ref{sim3}) obviously show that definitions $Q$ and $X$ terms are
exactly equal as expected, that is, the exchange of energy between
geometry and the scalar field in the noncommutative model are the
same as the exchange of energy between the matter and the scalar
field in the chameleon model. Also, as Eq. (\ref{conservPbeta})
indicates Eq. (\ref{nab}) in the chameleon model, one can
similarly have $\nabla_{\mu}{T^{[\ell]}}^{0\mu}=Q$ from Eq.
(\ref{conservPl}), where $T^{[\ell]}_{\mu\nu}$ can be interpreted
as an energy--momentum tensor associated to the noncommutativity
effect.

Furthermore, assuming the Hubble
parameter to be a constant during the inflation, relations
(\ref{cont1}), (\ref{scalefactor}) and (\ref{sim3}) actually yield
\begin{equation}
a^{-2}_{_{[\rm CH]}}\propto a^{-3(1+w)}_{_{[\rm CH]}} \qquad \quad \Longrightarrow \qquad\quad w= -1/3.
\end{equation}
That is, for a nearly constant $H$, the type of matter field is
constrained to be nearly a cosmic string--like during the
inflation. Also, these relations give $A=\pm\sqrt{\rho_{\rm
0}/(H_{\rm c}M^{3}_{\rm Pl})}$, where the positive sign is
acceptable for the inflationary expansion. In this case,
integrating result (\ref{sim4}) roughly gives
\begin{equation}\label{sim5}
\phi\sim \frac{2H_{\rm c}t}{\ell}+\phi_{\rm 0},
\end{equation}
where $\phi_{\rm 0}$ is an initial constant that, by
relations (\ref{cont1}) and (\ref{sim30}) with $w=-1/3$, is
$\phi_{\rm 0}\sim(2/\ell)\ln {a_{\rm 0}}_{_{[\rm CH]}}$. Solution
(\ref{sim5}) implies that the scalar field is roughly an
increasing function of time during the inflation.

On the other hand, using Eq. (\ref{friedman}) with relations
(\ref{sim1}) and (\ref{sim30}), and result (\ref{sim4}), we get
the chameleon potential to be
\begin{equation}\label{sim6}
V^{[\rm CH]}=\left[3M^{2}_{\rm
Pl}-\frac{9(1+w)^2}{2\,\ell^{2}}\right]H^{2}-\rho_{\rm 0},
\end{equation}
which is nearly constant during the inflation. Now, if one chooses
the potential of the chameleon model to be as a decreasing
function of $\phi$, where reduces nearly to a constant value while
the scalar field increases by time, then our proposed matching
procedure can be carried out. Such a choice is consistent with the
common typical chameleon potential that is usually used in the
literature, namely $V^{[\rm CH]}(\phi)=M^{4}exp(M^{n}/{\phi^{n}})$
or $V^{[\rm CH]}(\phi)=M^{4}+M^{n+4}/\phi^{n}$ with $n$ as a
positive integer describing the shape of the chameleon potential
($M$ is also a constant with the dimension of mass as an energy
scale that is often compared to cosmological--constant scale
($2.4$ meV) if the chameleon is to be relevant for the
present--day cosmic acceleration~\cite{wang2}).

On the other way, as the Hubble parameter is nearly constant
during the inflation, i.e., $\dot{H}\equiv-\epsilon H^{2}$ (where
$\epsilon$ is the slow--roll parameter), hence, by inserting $H$
from Eq. (\ref{fn1}) and $\dot{H}$ from Eq. (\ref{hdot}) into this definition
while using result (\ref{sim4}) and relations (\ref{sim3}) and
(\ref{sim30}), one gets the potential in this epoch to
be
\begin{equation}\label{sim7}
V^{[\rm
NC]}=\frac{9(1+w)^{2}}{\ell^{2}}\left(\frac{3}{2\,\epsilon}-\frac{1}{2}
\right)H^{2}+\frac{3(1+w)}{2}\frac{\rho_{\rm 0}}{\epsilon}
\end{equation}
with constraint
\begin{equation}\label{sim70}
V^{[\rm NC]}(\phi)>\frac{3(1+w)}{2}\frac{\rho_{\rm 0}}{\epsilon}.
\end{equation}
Hence, from equality of $V^{[\rm NC]}$ from solutions
(\ref{sim6}) and (\ref{sim7}), one can obtain the value of $\epsilon$ in terms
of the other parameters as
\begin{equation}\label{sim7000}
\epsilon=\frac{1+w}{2M^{2}_{\rm Pl}}\left[\frac{\rho_{\rm 0}}{H^{2}}+\frac{9(1+w)}{\ell^{2}}\right].
\end{equation}
Moreover, by differentiating potential (\ref{sim6}) while using
result (\ref{sim4}), we obtain
\begin{equation}\label{vprim}
V'_{\rm eff}=V'=-\frac{2\,\epsilon\,\ell}{3(1+w)}V^{[\rm NC]},
\end{equation}
that can also be achieved by inserting relation (\ref{sim1}) and
result (\ref{sim4}) into Eq. (\ref{phi}) while using relations
(\ref{sim30})  and (\ref{sim7000}). Hence, the price of our
proposal for matching the models is that the derivative of the
chameleon effective potential is a nearly negative constant value
in the inflationary epoch (for $w>-1$), however, it can be made to
be almost zero as being required in the chameleon models.

Obviously, Eq. (\ref{hdot2}) indicates that, onset of the
inflation where $\epsilon\ll1$ (that is, $\dot{H}\sim 0$), the
equation of state parameter should be $w\sim -1$, and thus, the
scalar field acts as a constant in this case. Hence, the
cosmological constant starts the extremely expansion of the
universe. Then, while the expansion is occurring, the matter field
being diluted and acts as a cosmic string--like. In this
situation, $V^{[\rm NC]}$ from solution (\ref{sim6}) with
constraint (\ref{sim70}) leads to a lower bound on the
noncommutative parameter, namely $\ell>\sqrt{2/3}/M_{\rm Pl}$,
where we have considered $\ell$ as a positive number. However,
$V^{[\rm NC]}$ from solutions (\ref{sim6}) and (\ref{sim7})
together, with constraint (\ref{sim70}), gives a more restricted
lower bound on $\ell$, namely
\begin{equation}\label{sim700}
\ell>\sqrt{\frac{2}{\epsilon}}\frac{1}{M_{\rm Pl}}.
\end{equation}
This constraint, via relation (\ref{sim1}) with $w=-1/3$, also
leads to a lower bound on the chameleon coupling constant as
\begin{equation}\label{betaepsilon}
\beta>\sqrt{\frac{1}{2\,\epsilon}}
\end{equation}
that, even considering $\epsilon $ to have an infinitesimal value
in the inflationary epoch (say, e.g., of the order  $10^{-5}$), it
still yields an acceptable estimate for $\beta $ consistent with
the recent experimental constraint obtained by Ref.~\cite{jaff},
namely $\beta < 3.7\times 10^2$. In turn, the recent experimental
constraint on $\beta $ also sets an upper bound on the value of
noncommutative parameter as $\ell < 7.2\times 10^2$ by relation
(\ref{sim1}) in an unit where $M_{\rm Pl} = 1$.

Now, with nearly constant Hubble parameter, i.e.,
$\dot{H}\equiv-\epsilon H^{2}$, although still with $w=-1/3$, by
manipulating result (\ref{sim4}) while using relation
(\ref{sim1}), we finally get an equation for the scalar field,
namely
\begin{equation}\label{phiii}
\ddot{\phi}+\frac{\beta\, \epsilon}{M_{\rm Pl}} \dot{\phi}^{2}=0.
\end{equation}
This concise equation indicates the behavior of the scalar field
with respect to time better than the roughly solution (\ref{sim5})
during the inflation (note that, the value of $\epsilon$ exactly
vanishes in solution (\ref{sim5})). The result of numerical
calculations of Eq. (\ref{phiii}) has been depicted in Fig.~(1)
for two different values of $\epsilon$ and the other constants
while considering the recent experimental constraint on $\beta $
and constraint (\ref{betaepsilon}) on the corresponding $\epsilon
$. The Figures confirm the roughly solution (\ref{sim5}) and
indicate that the scalar field linearly increases during the
inflation.
\begin{figure}[h]
\centerline{\includegraphics[scale=0.6]{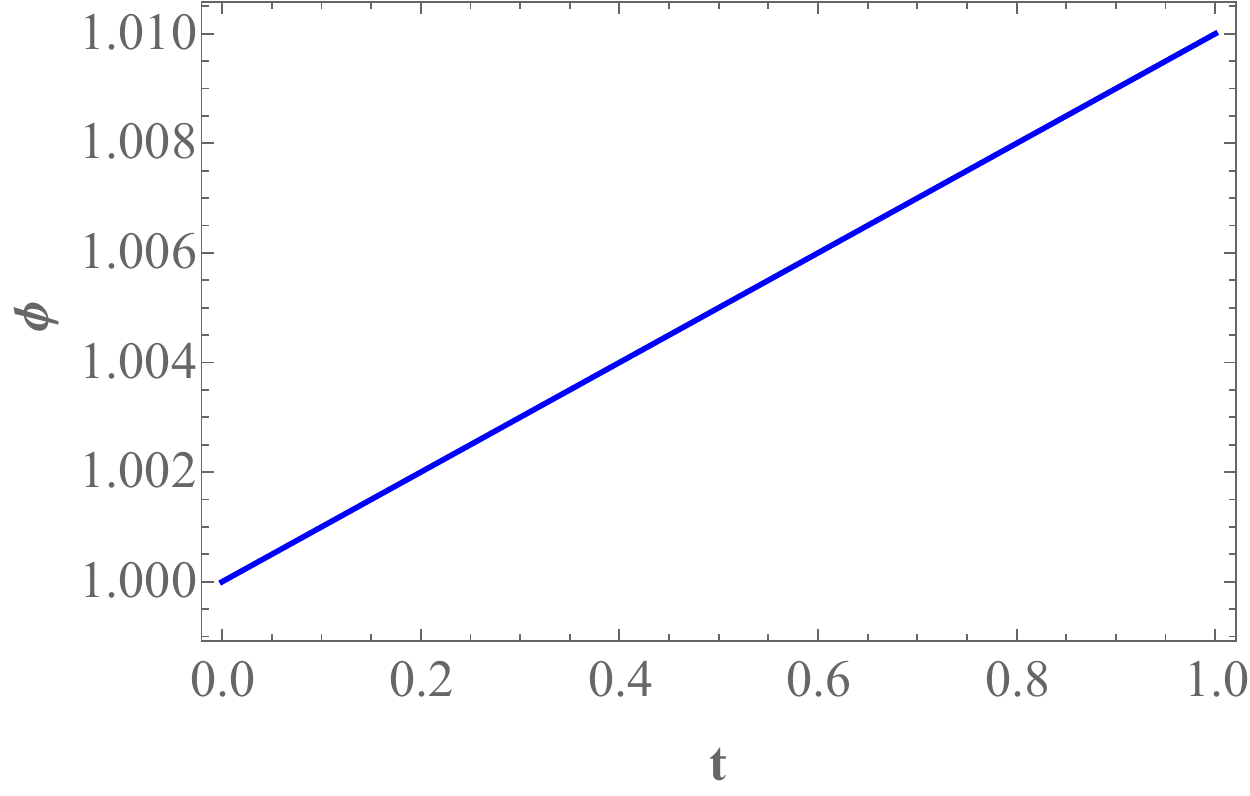}
\includegraphics[scale=0.6]{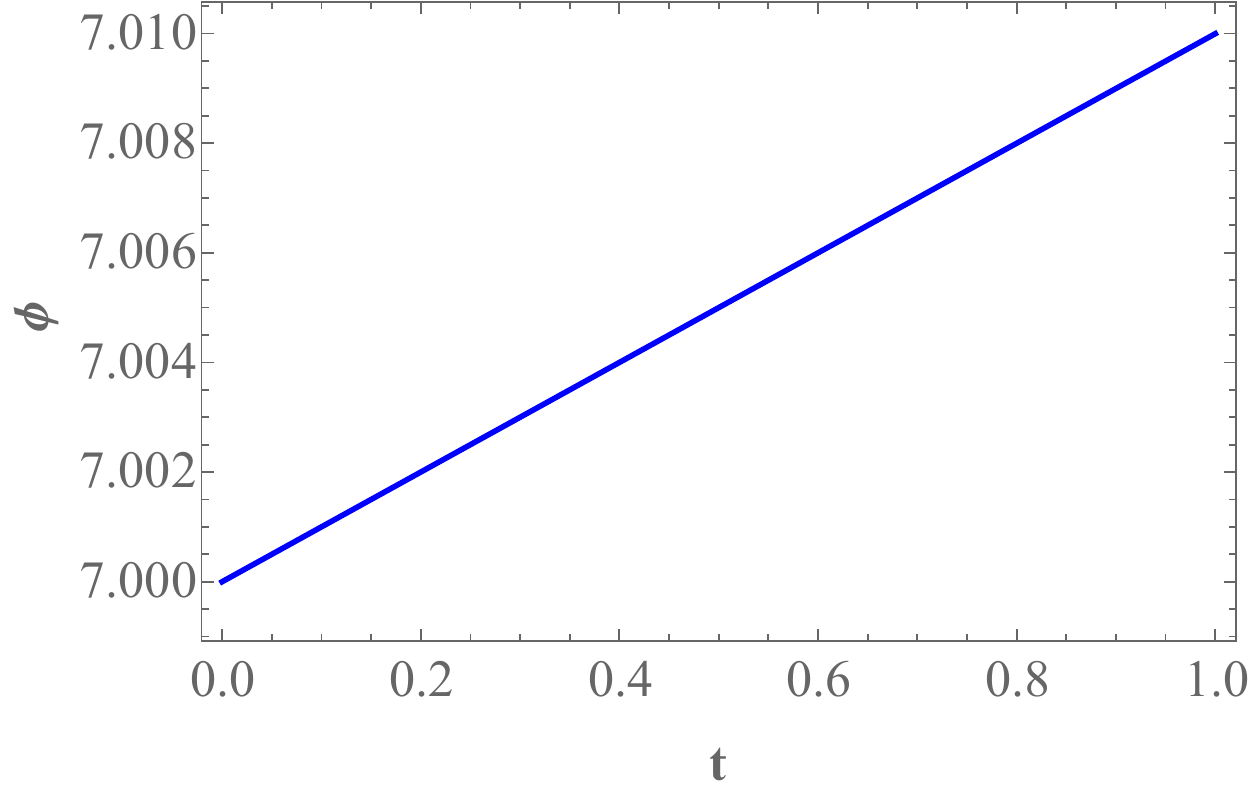}}
\caption{\label{}\small The figures show the time behavior of the
scalar field that have been plotted for the numerical values
$\beta=100$, $M_{\rm Pl}=1$ with initial value
$\dot{\phi_{0}}=0.01$, and (the left plot): $\epsilon=0.0001$ with
initial value $\phi_{0}= 1$, (the right plot): $\epsilon=0.001$
with initial value $\phi_{0}= 7$.}
\end{figure}

Our remaining task is to investigate the viability of the models
during the inflation. A successful inflationary model must resolve
the problems of standard cosmology. Hence, in order to solve the
horizon problem, which is more important than the flatness and
monopole problems (since by resolving this one, the other problems
will be solved automatically), inflation should last for a
sufficiently enough time in a way that its achieved number of
e--folding being at least around $50$ to $70$. In this respect,
the number of e--folding quantifies the amount of universe
expansion during the inflation, and is defined~\cite{wein2} to be
\begin{equation}\label{13}
N\equiv \int_{t_{b}}^{t_{e}} Hdt=\int_{\phi_{b}}^{\phi_{e}} \frac{H}{\dot{\phi}}d\phi,
\end{equation}
where the subscripts ``b" and ``e" refer to the beginning and the
end of inflation, respectively. By using result (\ref{sim4}) with
$w=-1/3$, one gets
\begin{equation}\label{14}
N=\frac{\ell}{2}\left(\phi_{e}-\phi_{b}\right).
\end{equation}
As the slow--roll condition is broken at the end of inflation,
thus the value of $\phi_{e}$ can be obtained via Eq. (\ref{phiii})
while setting the slow--roll parameter to be $\epsilon\sim1$.
Also, for the value of $\phi_{b}$, we use relation~\cite{ruth}
\begin{equation}\label{i1}
r\sim 16\,\epsilon|_{\phi=\phi_{b}},
\end{equation}
where the parameter $r$ is the ratio of the tensor perturbation
amplitude to the scalar perturbation amplitude with an upper bound
$r<0.11$ in $95\%$ confidence level stated in the Planck
measurements~\cite{p13}--\cite{p152}. Then, by tuning the free
parameters and also the initial values for the scalar field and
its derivative, one can achieve an appropriate number of
e--folding.

Therefore, through the analogy of the noncommutative standard and
chameleon models, it has been shown that the chameleon model can
be a successful one during the inflationary epoch. Also, we
conclude that one can consider a single scalar field being
responsible for both roles of the inflaton and the chameleon.

\section{Conclusions}
\indent

In this work, we have indicated that there is a correspondence
between the chameleon model (where the chameleon scalar field
non--minimally couples with the matter field) and the
noncommutative standard model (in which the inflaton scalar field
minimally couples with gravity). On the other hand, in another
work~\cite{RSFM}, we have shown that a noncommutative inflationary
model, in the presence of a particular type of dynamical
deformation between the canonical momenta of the scale factor and
of the scalar field, is a successful model during the inflation.
Now, through the mentioned correspondence procedure of this work,
we have indicated that there is a relevance between the
noncommutative parameter and the chameleon coupling strength. Such
a correspondence, in turn, presents that noncommutative effects
act as a source term in the Klein--Gordon equation in a similar
manner to the matter density in a chameleon theory, wherein the
matter density of the environment in the chameleon model is
related to influence of geometry in the noncommutative standard
model. In fact, the results of the work, as a complementary to our
previous work~\cite{saba}, provide a possibility to take into
account the noncommutative standard model instead of the chameleon
model during the inflation. However, we should emphasize that the
specific choice of the dynamical deformation (that has been
assumed between the canonical momenta of the scale factor and of
the scalar field) is very constructive to reach in such a
relevance.

Moreover, this correspondence has represented that the type of the
matter field in the chameleon model is constrained to be nearly a
cosmic string--like during the inflation. The cosmic string, first
introduced by Kibble in the 1970s, is a kind of hypothetical
object associated to topological defects. It is claimed to be
formed during the very early universe~\cite{cosmic0,cosmic1} and
has been predicted in some field theory models. Also, its
formation in the context of string theory has attracted some
attentions (see, e.g., Ref.~\cite{cosmic2} and references therein)
and few efforts have recently been devoted to search for evidence
of its existence.

Nevertheless, through the obtained correspondence, the
noncommutative standard model provides a viable chameleonic
inflationary model as we proposed in Ref.~\cite{saba}. In fact,
there exist two scenarios for the evolution of the universe. In
the first one, it can be described by the noncommutative standard
model where the influence of geometry drives the universe
expansion, whereas in the second scenario, one considers the
evolution of the universe via the chameleon model. In this regard,
at the beginning of the inflation, when the energy density of the
universe is extremely high, the chameleon field acquires a very
large mass that makes the effect of its coupling suppressed. Thus,
the matter density acts as the cosmological constant and starts
the inflation. However, during the inflation, due to extreme
expansion of the universe, the matter field being diluted and acts
as a cosmic string--like. Therefore, at the beginning of the
inflation, the cosmological constant drives the inflation and
then, the scalar field plays the role of inflaton. Thus, the
evolution of the universe being described by one single scalar
field during the inflation that plays the role of inflaton in the
very early universe and then, acts as a chameleon field.
Furthermore, our proposed correspondence procedure has set some
constraints on the noncommutative parameter and the chameleon
coupling constant as well as nearly specifying functions of the
scalar field and its potential. Through matching the models, we
have also obtained a nearly negative constant value for the
derivative of potential in the inflationary epoch that can be set
to be as close to zero as required to be consistent with the
chameleon model conditions.

By the way, the investigation of the chameleon model, when the
universe reheats after inflation, can also be of much interest
that we propose to study it in a subsequent work.
\section*{Acknowledgement}
\indent

We thank the Research Council of Shahid Beheshti University for
financial support.

%
\end{document}